\documentstyle[12pt]{article}
\begin{document}
\begin{center}
{\textmd{\bf LINEAR THERMAL INSTABILITY AND FORMATION OF CLUMPY
GAS CLOUDS INCLUDING THE AMBIPOLAR DIFFUSION}}
~\\
~\\
{\textsc{mohsen nejad-asghar} \footnote {E-mail: nejad6601@wali.um.ac.ir}}
\& {\textsc{jamshid ghanbari} \footnote {E-mail: ghanbari@ferdowsi.um.ac.ir}}\\
~\\
~\\
~\\
Department of physics, School of Sciences,Ferdowsi University of Mashhad, Mashhad, Iran
~\\
~\\
\end{center}
\date {}
~\\
~\\
\newpage
\begin{abstract}
Thermal instability is one of the most important processes in the
formation of clumpy substructure in magnetic molecular clouds. On
the other hand, ambipolar diffusion, or ion-neutral friction, has
long been thought to be an important energy dissipation mechanism
in these clouds. Thus, we would interested to investigate the
effect of ambipolar diffusion on the thermal instability and
formation of clumps in the magnetic molecular clouds. For this
purpose, in the first step, we turn our attention to the linear
perturbation stage. In this way, we obtain a non-dimensional
characteristic equation which reduces to the prior characteristic
equation in the absence of the magnetic field and ambipolar
diffusion. With numerical manipulation of this characteristic
equation, we conclude that there are solutions where the thermal
instability allows compression along the magnetic field but not
perpendicular to it. We infer that this aspect might be an
evidence in formation of observed disc-like (oblate) clumps in
magnetic molecular clouds.
\end{abstract}
~\\
~\\
{\textbf{key words:} globular clusters: general- instabilities-
diffusion- ISM: clouds}
\newpage
\section{Introduction}
An almost thorough analysis of linear stage of thermal
instability was given in a well-known paper by Field(1965). He
showed that thermal instability can lead to the rapid growth of
density perturbations from infinitesimal $\delta\rho$ to nonlinear
amplitudes on a cooling time-scale, $\tau_c$, in which for typical
conditions in the interstellar medium (ISM) is short compared to
the dynamical time-scale $\tau_d$. Of the purely thermal modes,
the most relevant for the ISM is the isobaric mode which has been
discussed in terms of the formation of distinct phases of the ISM
(Field, Goldsmith \& Habing 1969) and the formation of protostars
in a cooling ISM (Schwarz, McCray \& Stein 1972). The isentropic
mode has also found application in the ISM and has been discussed
with regard to the amplification of acoustic waves in a warm ISM
($30<T<90\;K$) (Flannery \& Press 1979). A more detailed
investigation of the growth of condensations in cooling regions
has been presented by Balbus(1986,1991) who examined the effect
of magnetic field.
\\High resolution studies of the magnetic molecular clouds,
reveal that they have internal structures on all scales and are
typically clumpy or filamentary (Falgarone, Puget \& P\'{e}rault
1992, Langer et al 1995), with prolate and oblate (disc-like)
clumps (Ryden 1996). Gammie et al (2003) have recently studied the
three dimensional analogs of clumps. They have concluded that
nearly $90\%$ of the clumps are prolate and $10\%$ of them are
oblate. The origin and shape of these clumps is a disputable
issue. Thermal instability and turbulence may be two responsible
parameters.
\\In molecular clouds, the dispersion velocity inferred from
molecular line width is often larger than the gas sound speed
inferred from transition temperatures (Solomon et al 1987). MHD
turbulence may be responsible for the stirring of these clouds
(Arons \& Max 1975). because of these turbulent motions,
molecular clouds must be transient structures, and are probably
dispersed after not much more than $\sim 10^7 yr$ (Larson 1981).
Since cooling time-scale of molecular clouds is approximately
$\sim 10^3-10^4 yr$ (Gilden 1984), thermal instability may be a
coordinated trigger mechanism for clump formation. Turbulence, in
the second stage, can deform these small-scale clumps in shape
and orient them relative to the magnetic fields.
\\Gilden(1984) calculated the net cooling function for molecular
clouds and found that in environments where CO cooling dominates,
molecular gas may be thermally unstable. He suggested that
thermal instability may be an important source of small-scale
clumps in fully molecular clouds. Burkert \& Lin(2000) have
recently proposed that clumpyness in cold clouds arises naturally
from their formation through a cooling instability which acts on
time-scales that can be much shorter than the dynamical
time-scale of the cloud. Afterward, Gomez-Pelaez \&
Moreno-Insertis(2002) have investigated the effect of
self-gravity and thermal conduction on a cooling and expanding
medium. They classified importance of various physical processes
including self-gravity, background expansion, cooling, and
thermal conduction according to their relative time-scales.
\\On the other hand, observations establish that magnetic fields
play an important role in shaping the structure and dynamics of
molecular clouds. Especially, ambipolar drift, or ion-neutral
friction, has long been thought to be an important energy
dissipation mechanism in magnetic molecular clouds (Scalo 1977).
If $\lambda$ represents the characteristic dimension over which
the magnetic field varies (the wavelength of perturbation), the
time-scale of ambipolar diffusion in a typical molecular cloud
may be approximated as $\tau_{AD}\sim 10^8 \lambda_{(PC)}^2 yr$
(Shu 1991). We expect that for a critical wavelength
($\sim0.01pc$) which $\tau_{AD}\sim\tau_c$, ambipolar diffusion
may be important. Thus, in view of this importance, we would
interested to investigate the effect of it on thermal instability
and formation of small-scale clumps in the magnetic molecular
clouds. We suggest that shape of oblate clumps is formed from
their early stage of evolution, via thermal instability.
\\For this purpose, in the first step, we turn our attention to the
linear stage and neglect the effect of self-gravity and background
expansion or contraction. Section II of this paper develops the
theory of linear thermal instability in the presence of ambipolar
diffusion. In {\S III}, we obtain a non-dimensional
characteristic equation which can reduce to the Field's
characteristic equation in the absence of ambipolar diffusion.
Then, we discuss about the domains of stability (or instability)
of this characteristic equation. We find a critical wavelength
which the effect of ambipolar diffusion is very important and
small-scale disc-like clumps can be formed. Finally, a conclusion
is given in {\S IV}.
\section{The Linearized Equations}
A molecular cloud gas includes neutral atoms and molecules, atomic
and molecular ions, and electrons, which are the primary current
carriers. Since significant charge separation can not be
sustained on the time-scale of interest, so the electrons and
ions move together.
\\In principle, the ion velocity $\textbf{\textit{v}}_i$ and the
neutral velocity $\textbf{\textit{v}}_n$ should be determined by
solving separate fluid equations for  these species (Draine 1986),
including their coupling by collision processes. But, in our
interested time-scale of cooling ($10^3-10^4 yr$, Gilden 1984),
two fluids of ion and neutral are well coupled by together, thus
we can use the basic equations as follows (Shu 1991)
\begin{equation}
\frac{d\rho}{dt} + \rho\nabla\cdot\textbf{\textit{v}}=0
\end{equation}
\begin{equation}
\rho\frac{d\textbf{\textit{v}}}{dt} + \nabla p +
\nabla(\frac{B^2}{8\pi}) -
(\textbf{B}\cdot\nabla)\frac{\textbf{B}}{4\pi}=0
\end{equation}
\begin{equation}
\frac{1}{\gamma-1}\frac{dp}{dt} -
\frac{\gamma}{\gamma-1}\frac{p}{\rho}\frac{d{\rho}}{dt} +
\rho\Omega-\nabla\cdot(K\nabla T)=0
\end{equation}
\begin{equation}
\frac{d\textbf{B}}{dt}+\textbf{B}(\nabla\cdot\textbf{\textit{v}})-(\textbf{B}\cdot\nabla)\textbf{\textit{v}}=
\nabla\times\{\frac{\textbf{B}}{4\pi\eta\epsilon\rho^{1+\nu}}\times
[\textbf{B}\times(\nabla\times\textbf{B})]\}
\end{equation}
\begin{equation}
p-\frac{R}{\mu}\rho T=0
\end{equation}
where $d/dt=\partial/\partial t+\textbf{\textit{v}}\cdot\nabla$
is the Lagrangian time derivative and $K$ is the coefficient of
thermal conduction. $\gamma$ is the polytropic index of the ideal
gas, $\mu$ is the mean atomic mass per particle, $R$ is the
universal gas constant, and
$\eta=\frac{<v_{in}\sigma_{in}>}{m_i+m_n}$ where $v_{in}$ is the
ion-neutral relative velocity with impinging cross section
$\sigma_{in}$. In writing the above equations, we used the
relation $\rho_i=\epsilon\rho_n^\nu$ between ion density and
neutral density (Nakano 1980). Since, the ion density is much
less than the neutral density, in a good approximation we estimate
\begin{equation}
\rho=\rho_n+\rho_i\approx\rho_n.
\end{equation}
$\Omega(\rho,T)$ is the net cooling function
($erg.sec^{-1}.gr^{-1}$) as follows
\begin{equation}
\Omega(\rho,T)=\Lambda(\rho,T)-\Gamma_{tot}
\end{equation}
where $\Gamma_{tot}$ is the total heating rate and
$\Lambda(\rho,T)$ is the cooling rate which can be written as
(Goldsmith \& Langer 1978)
\begin{equation}
\Lambda(\rho,T)=\Lambda_0 \rho^\delta T^\beta
\end{equation}
where $\Lambda_0$, $\delta$, and $\beta$ are constants. The range
of $\beta$ is $1.4$ to $2.9$. The constant $\delta$ is greater
than zero for optically thin case and less than zero for
optically thick case. Models of molecular clouds identify several
different heating mechanisms; cosmic rays, $H_2$ formation, $H_2$
dissociation, grain photoelectrons, collisions with warm dust,
gravitational contraction, and ambipolar diffusion. The rate for
these processes are largely unknown. Sticking efficiency on
grains, grain composition and lattice structures, cosmic ray
spectra and flux, efficiency of cosmic ray penetration into
clouds, magnetic field strengths and geometry, and the fractional
ionization are a few disputable parameters. In this paper we
consider the heating rates of cosmic rays, $H_2$ formation, $H_2$
dissociation, grain photoelectrons, and collisions with warm dust
as a constant $\Gamma_0$ (Glassgold \& Langer 1974, Goldsmith \&
Langer 1978). Also, we disregard gravitational heating rate,
since we neglect self-gravity and background contraction. The
heating of the gas by magnetic ion-neutral slip is discussed in
detail by Scalo(1977); our simple estimate of this heating rate
is as follows
\begin{equation}
\Gamma_{AD}=\eta\epsilon\rho^\nu v_d^2
\end{equation}
where $v_d$ is the drift velocity of the ions relative to the
neutral. With above explanations/approximations and with
introduction $\Gamma_0'\equiv\eta\epsilon v_d^2$, in a good
manner we choose the net cooling function as follows
\begin{equation}
\Omega(\rho,T)=\Lambda_0\rho^\delta
T^\beta-\Gamma_0-\Gamma_0'\rho^\nu.
\end{equation}
In the local homogeneous equilibrium state, we have
$\rho=\rho_0,T=T_0,p=p_0,\textbf{B}=\textbf{B}_0,\textbf{\textit{v}}=0$,
and $\Omega(\rho_0,T_0)=0$. We assume perturbations of the form
\begin{equation}
A(\textbf{\textit{r}},t)=A_1
exp(ht+i\textbf{\textit{k}}\cdot\textbf{\textit{r}}).
\end{equation}
Then the linearized equations are
\begin{equation}
h\rho_1+i\rho_0\textbf{\textit{k}}\cdot\textbf{\textit{v}}_1=0
\end{equation}
\begin{equation}
h\rho_0\textbf{\textit{v}}_1+i\textbf{\textit{k}}p_1+i(\textbf{B}_0\cdot\textbf{B}_1)
\frac{\textbf{\textit{k}}}{4\pi}-
i(\textbf{\textit{k}}\cdot\textbf{B}_0)\frac{\textbf{B}_1}{4\pi}=0
\end{equation}
\begin{equation}
\frac{h}{\gamma-1}p_1-\frac{h\gamma p_0}{(\gamma-1)\rho_0}\rho_1+
\rho_0\Omega_\rho\rho_1+\rho_0\Omega_TT_1+Kk^2T_1=0
\end{equation}
\begin{equation}
h\textbf{B}_1+i\textbf{B}_0(\textbf{\textit{k}}\cdot\textbf{\textit{v}}_1)-i(\textbf{\textit{k}}\cdot\textbf{B}_0)\textbf{\textit{v}}_1=
i\textbf{\textit{k}}\times\{\frac{\textbf{B}_0}{4\pi\eta\epsilon\rho_0^{1+\nu}}\times[\textbf{B}_0\times
(i\textbf{\textit{k}}\times\textbf{B}_1)]\}
\end{equation}
\begin{equation}
\frac{p_1}{p_0}-\frac{\rho_1}{\rho_0}-\frac{T_1}{T_0}=0
\end{equation}
where $\Omega_\rho\equiv(\partial\Omega/\partial\rho)_T$ and
$\Omega_T\equiv(\partial\Omega/\partial T)_\rho$ are evaluated
for the equilibrium state.\\
We introduce the coordinate system
$\textbf{\textit{u}}_x,\textbf{\textit{u}}_y,\textbf{\textit{u}}_z$
specified by
\begin{equation}
\textbf{\textit{u}}_z=\frac{\textbf{B}_0}{B_0}\quad,\quad\textbf{\textit{u}}_y=\frac{\textbf{B}_0\times\textbf{\textit{k}}}
{|\textbf{B}_0\times\textbf{\textit{k}}|}\quad,\quad\textbf{\textit{u}}_x=\textbf{\textit{u}}_y\times\textbf{\textit{u}}_z.
\end{equation}
Equations (13) and (15) may be used to uncouple $v_{1y}$- the
perturbed velocity in the plane perpendicular to both
$\textbf{B}_0$ and $\textbf{\textit{k}}$- from the reminder of
the problem. Disturbances perpendicular to the
$(\textbf{B}_0-\textbf{\textit{k}})$-plane have a solution of the
form
\begin{equation}
h=-\frac{(k_\parallel a)^2}{2\eta\epsilon\rho_0^\nu}\pm i
k_\parallel a[1-(\frac{k_\parallel
a}{2\eta\epsilon\rho_0^\nu})^2]^{\frac{1}{2}}
\end{equation}
where $a$ is the Alfv\'{e}n velocity and
$k_\parallel=\textbf{\textit{k}}\cdot\textbf{\textit{u}}_z$.
Thus, in long wavelength perturbations that $k_\parallel
a<2\eta\epsilon\rho_0^\nu$, waves are damped with ion-neutral
friction and in short wavelength perturbations that $k_\parallel
a>2\eta\epsilon\rho_0^\nu$, Alfv\'{e}n waves can not be existed.
\\The motion in the other modes are constrained to the $x-z$-plane,
and are governed by the characteristic equation,\\
$\textbf{h}^5+(c_sk_T+h')\textbf{h}^4+[k^2(c_s^2+a^2)+h'k_Tc_s]\textbf{h}^3+
\{k^2[c_s^3(k_T-k_\rho)/\gamma+a^2c_sk_T]+c_s^2h'k^2\}\textbf{h}^2$
\begin{equation}
+[c_s^2a^2k^4\cos^2\theta+k^2c_s^3h'(k_T-k_\rho)/\gamma]\textbf{h}+
(k_T-k_\rho)c_s^3a^2k^4\cos^2\theta/\gamma=0
\end{equation}
where $c_s$ is the Laplacian speed of sound, $\theta$ is the
angle between $\textbf{\textit{k}}$ and $\textbf{B}_0$, and
\begin{equation}
k_\rho=\frac{\mu(\gamma-1)\rho_0\Omega_\rho}{Rc_sT_0}\quad,\quad
k_T=\frac{\mu(\gamma-1)\Omega_T}{Rc_s}+\frac{\mu(\gamma-1)K}{Rc_s\rho_0}k^2
\end{equation}
are wavenumbers of sound waves whose angular frequencies are
numerically equal to the growth rates of isothermal and isochoric
perturbations, respectively. $h'\equiv
k^2a^2/\eta\epsilon\rho_0^\nu$ is the effect of ion-neutral slip.
If we neglect the effect of the magnetic field and ambipolar
diffusion ($a=0,h'=0$), the characteristic equation (19) reduces
to the equation (15) of Field(1965).
\section{Domains of Stability}
By introducing the non-dimensional quantities,
\begin{equation}
y\equiv\frac{h}{kc_s}\quad,\quad\sigma_\rho\equiv\frac{k_\rho}{k}\quad,\quad
\sigma_T\equiv\frac{k_T}{k}
\quad,\quad\alpha\equiv(\frac{a}{c_s})^2\quad,\quad
D\equiv\frac{h'}{kc_s}
\end{equation}
we can write the characteristic equation in the following form,\\
$\textbf{y}^5+(\sigma_T+D)\textbf{y}^4+(1+\alpha+\sigma_TD)\textbf{y}^3
+[\gamma^{-1}(\sigma_T-\sigma_\rho)+\alpha\sigma_T+D]\textbf{y}^2$
\begin{equation}
+[\alpha\cos^2\theta+\gamma^{-1}(\sigma_T-\sigma_\rho)D]\textbf{y}
+\gamma^{-1}\alpha(\sigma_T-\sigma_\rho)\cos^2\theta=0
\end{equation}
so that for each $\theta$ we have four free parameters that
consist of $\sigma_T$, $\sigma_\rho$, $\alpha$, and $D$. We want
to study the effect of ambipolar diffusion on stable region of the
$\sigma_T-\sigma_\rho$ plane. For this purpose, we use the
Laguerre's method for finding the roots of the characteristic
equation.
\\First we consider the problem in the absence of the magnetic
field ($\alpha=0,D=0$). The stable region of this case is shown
in Fig.~1. This result had been derived by Field(1965). Now, we
are interested to consider the effect of the magnetic field. For
this purpose, we must consider the ambipolar diffusion because of
small ion density in magnetic molecular clouds. In this case, we
must break a lance to the complete characteristic equation (22)
for different values of $\sigma_T$, $\sigma_\rho$, $\alpha$, $D$,
and $\theta$. The stable regions of these typical instances are
shown in Fig.~2. In this case, the line $OA$ in Fig.~1 is
unchanged, corresponding to the breaking of the magnetic pressure
for the reason of ion-neutral slipping and smallness of ions.
But, the line $OB$ in Fig.~1 is brought down, corresponding to the
dissipating ion-neutral slip heating during the compression phase
of the wave. Thus, ambipolar diffusion can stabilize the medium
so that it's maximum effect is occurred at $\theta=\pi/2$.
\\Inserting the net cooling function, Equ.(10), into the
definitions of $\sigma_\rho$ and $\sigma_T$, we get
\begin{equation}
\sigma_\rho=\frac{\mu(\gamma-1)}{R c_s k
T_0}\Lambda(\rho_0,T_0)(\delta-\nu\xi)
\end{equation}
\begin{equation}
\sigma_T=\frac{\mu(\gamma-1)}{R c_s k
T_0}\beta\Lambda(\rho_0,T_0)[1-(\frac{\lambda_0}{\lambda})^2]
\end{equation}
where $\xi$ is the ratio of ambipolar diffusion heating rate to
the cooling rate as
\begin{equation}
\xi\equiv\frac{\Lambda(\rho_0,T_0)}{\Gamma_0'\rho_0^\nu}
\end{equation}
and $\lambda_0$ is a defined wavelength as follows
\begin{equation}
\lambda_0\equiv
2\pi\sqrt{\frac{KT_0}{\beta\rho_0\Lambda(\rho_0,T_0)}}.
\end{equation}
We separate two cases as follows
\begin{enumerate}
\item $\delta>\xi\nu$ which $\sigma_\rho>0$ that is upwards of
the $\sigma_T-\sigma_\rho$ plane. In this case, the medium is
optically thin. As shown in Fig.~3(a), two regions of the
$\sigma_T-\sigma_\rho$ plane are separated by a critical
wavelength
\begin{equation}
\lambda_{c1}\equiv\frac{\lambda_0}{\sqrt{1-\frac{\delta-\xi\nu}{\beta}}}.
\end{equation}
If the wavelength of perturbation is greater than this critical
value, medium is stable. If $\lambda<\lambda_{c1}$, the magnetic
molecular cloud is unstable and a spherical clump can be formed.
\item $\delta<\xi\nu$ which $\sigma_\rho<0$ that is downwards of the $\sigma_T-\sigma_\rho$
plane.The optically thick molecular clouds set in this case. As
shown in the Fig.~3(b), three regions of the
$\sigma_T-\sigma_\rho$ plane are separated by two critical
wavelengths
\begin{equation}
\lambda_{c2}\equiv\frac{\lambda_0}{\sqrt{1+\frac{\xi\nu-\delta}{\beta}}},\quad
\lambda_{c3}\equiv\frac{\lambda_0}{\sqrt{1-\frac{3}{2}\frac{\xi\nu-\delta}{\beta}}}.
\end{equation}
If the wavelength of perturbation is greater than $\lambda_{c3}$,
the medium is stable. If $\lambda<\lambda_{c2}$, the molecular
cloud is unstable and a spherical clump can be formed. Depend on
the values of $D$ and $\alpha$, we have semi-stable regions
between $\lambda_{c2}$ and $\lambda_{c3}$, which the thermal
instability allows compression along the magnetic field but not
perpendicular to it.
\end{enumerate}
The increased area of semi-stable region (shaded areas of Fig.~2)
as a function of $D$ for three typical values of $\alpha$ and for
$\theta=\pi/2$ is plotted in Fig.~4. According to this figure,
maximum of stability is occurred at a typical $D_m$. Thus, we may
define a critical wavenumber
\begin{equation}
k_c=\frac{\eta\epsilon\rho_0^\nu c_s}{a^2}D_m.
\end{equation}
At small wavelengths which $k\gg k_c$, ambipolar diffusion can
break the effect of the magnetic field on the whole matter, thus
the value of $A_S$ is zero. On the other hand, at very large
wavelengths which $k\ll k_c$, ambipolar diffusion time-scale is
very greater than the cooling time-scale, therefore we can neglect
its effect , thus the value of $A_S$ must be zero, too.
\section{Conclusion}
We have carried out a systematic analysis of the linear thermal
instability of a locally uniform magnetic molecular cloud which,
in the perturbed state, is undergoing ambipolar diffusion.
Although thermal instability proceeds faster than dynamical
processes such as turbulence, its growth rate is determined by
the local cooling rate. We choose a simple parametric net cooling
function and discuss about its different parameters for
unstability and clump formation. The small perturbation problem
yields a complete characteristic equation that in the absence of
the magnetic field and ambipolar diffusion, reduces to the prior
results of the linear thermal instability. We have used the
Laguerre's method for finding the roots of this characteristic
equation.
\\The stable region by neglecting the magnetic field is shown
in Fig.~1, while, Fig.~2 displays the stability region for typical
values of the magnetic field ($\alpha$) and the ambipolar
diffusion strength (D). Comparison of these figures indicate that
ambipolar diffusion can stabilize the medium, in this manner that
its maximum stabilization is occurred perpendicular to the
magnetic field ($\theta=\pi/2$). Thus, including the magnetic
field and considering the ambipolar diffusion, divides the
$\sigma_T-\sigma_\rho$ plane in three regions: stable region,
unstable region, and semi-stable region.
\\By inserting the parametric general form of net cooling function
into the definitions of $\sigma_T$ and $\sigma_\rho$, we find
critical wavelengths which divide different cases of stability,
instability, and semi-stability of the $\sigma_T-\sigma_\rho$
plane according to the wavelength of perturbation. If the physical
parameters of the molecular cloud (or the wavelength of
perturbation) settle on the unstable region of Fig.~3, a
spherical clump must be formed. On the other hand, if its
parameters or the wavelength of perturbation settle on the
semi-stable region, thermal instability allows compression along
the local magnetic field but not perpendicular to it. Therefore,
including the magnetic field and ambipolar diffusion may be an
evidence in formation of small-scale disc-like clumps in magnetic
molecular clouds.
\\Since ambipolar diffusion time-scale depends on the
wavelength of perturbation, we find a critical wavenumber which
the effect of ambipolar diffusion for stabilizing the medium is
very important.
\\We have assumed a uniform background. In spite of this, our
results are applicable locally in a non-uniform background if the
perturbation wavelength is much less than the macroscopic
variation length of the unperturbed quantities.
\\We neglected the interaction and merging of the clumps. These
processes become important for the subsequent evolution. We also
neglected the effect of self-gravity and contraction or expansion
of the background. They will be considered in the subsequent
papers.

\newpage
\input{epsf}
\epsfxsize=4in \epsfysize=6in
\begin{figure}
\centerline{\epsffile{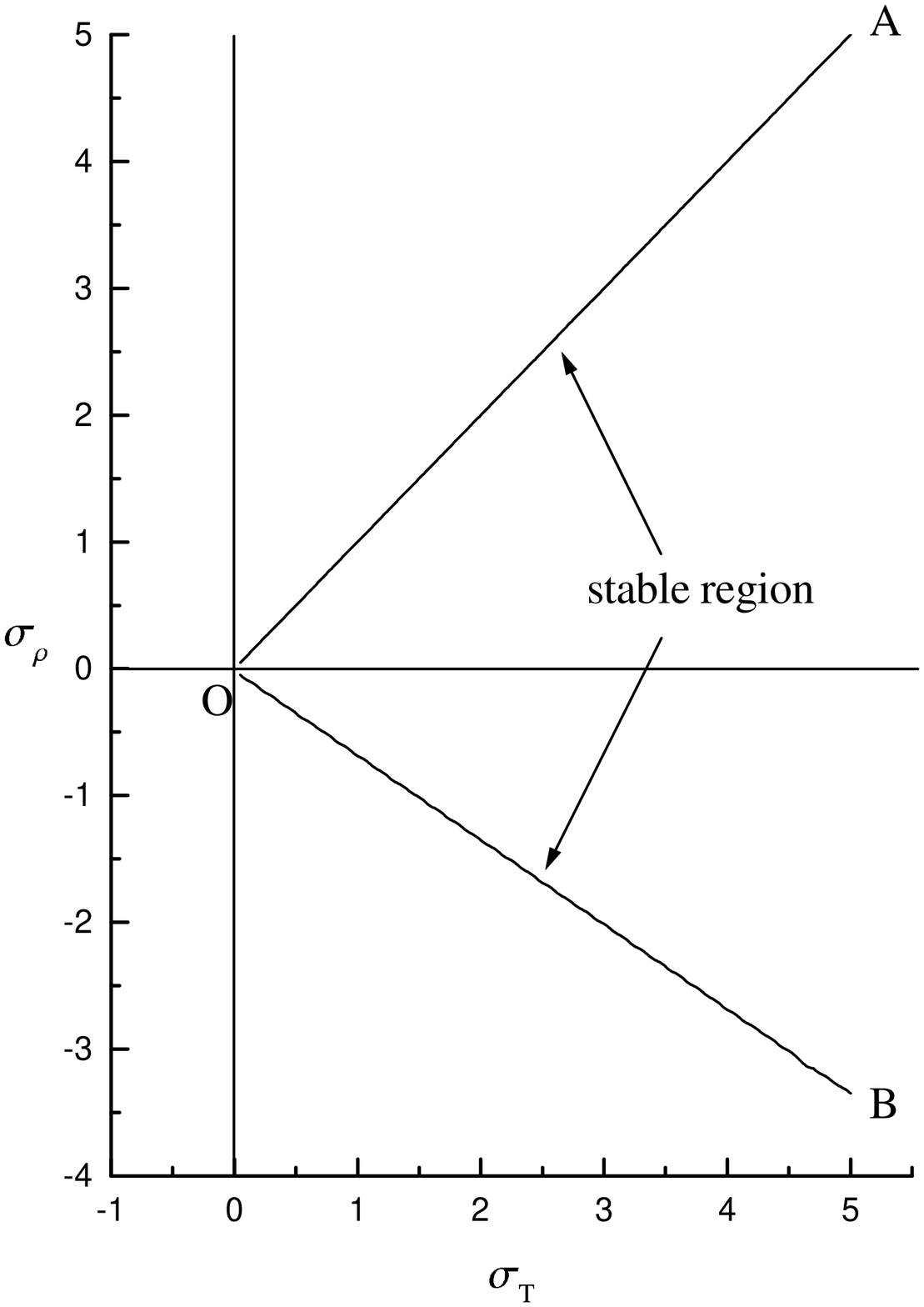}}
  \caption[]{Region of stability in the $\sigma_T-\sigma_\rho$ plane
  in the absence of the magnetic field ($\alpha=0,D=0$).}
\end{figure}
\newpage
\input{epsf}
\begin{figure}
 \centerline{{\epsfxsize=1.7in\epsfysize=2.2in\epsffile{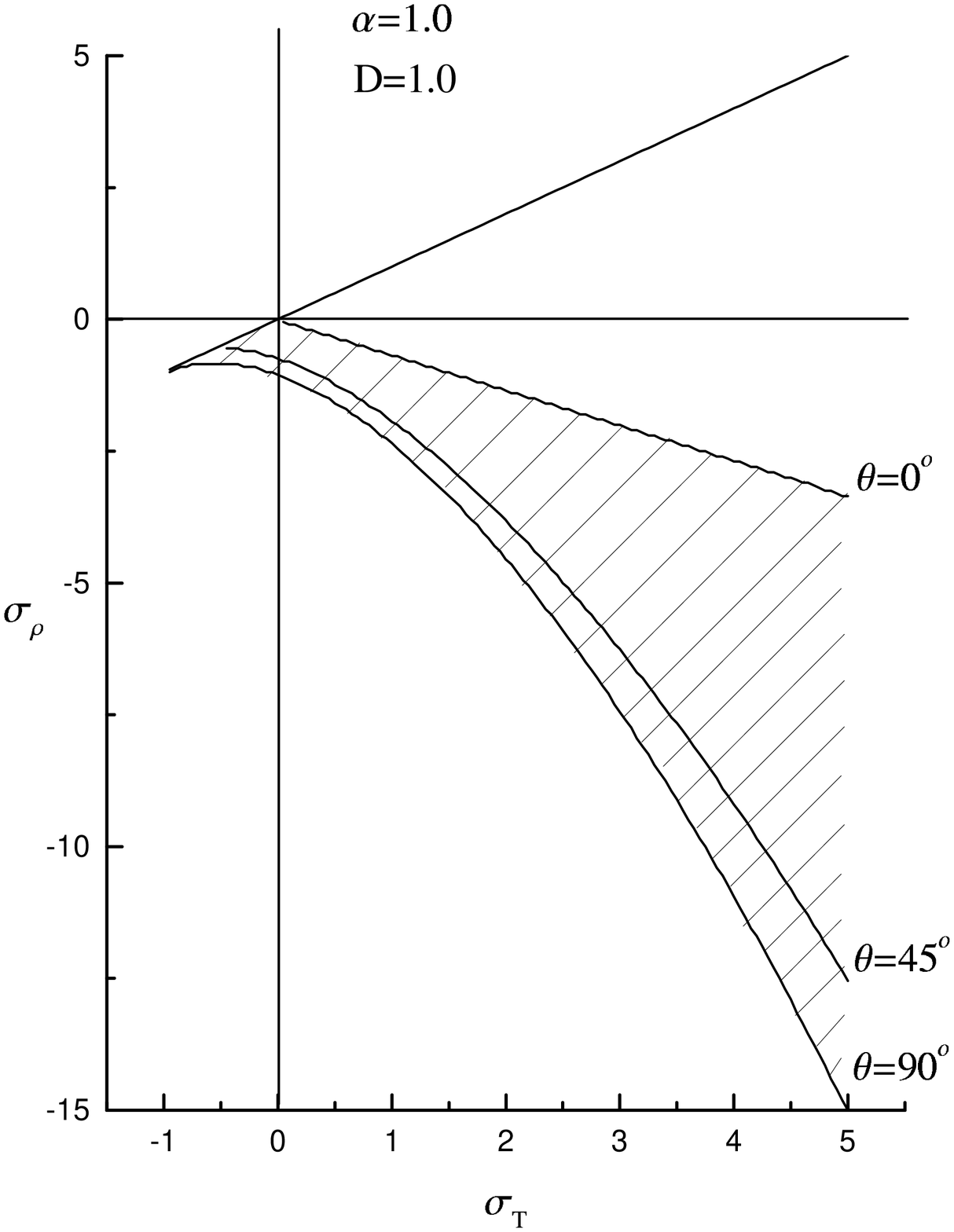}}
   {\epsfxsize=1.7in \epsfysize=2.2in\epsffile{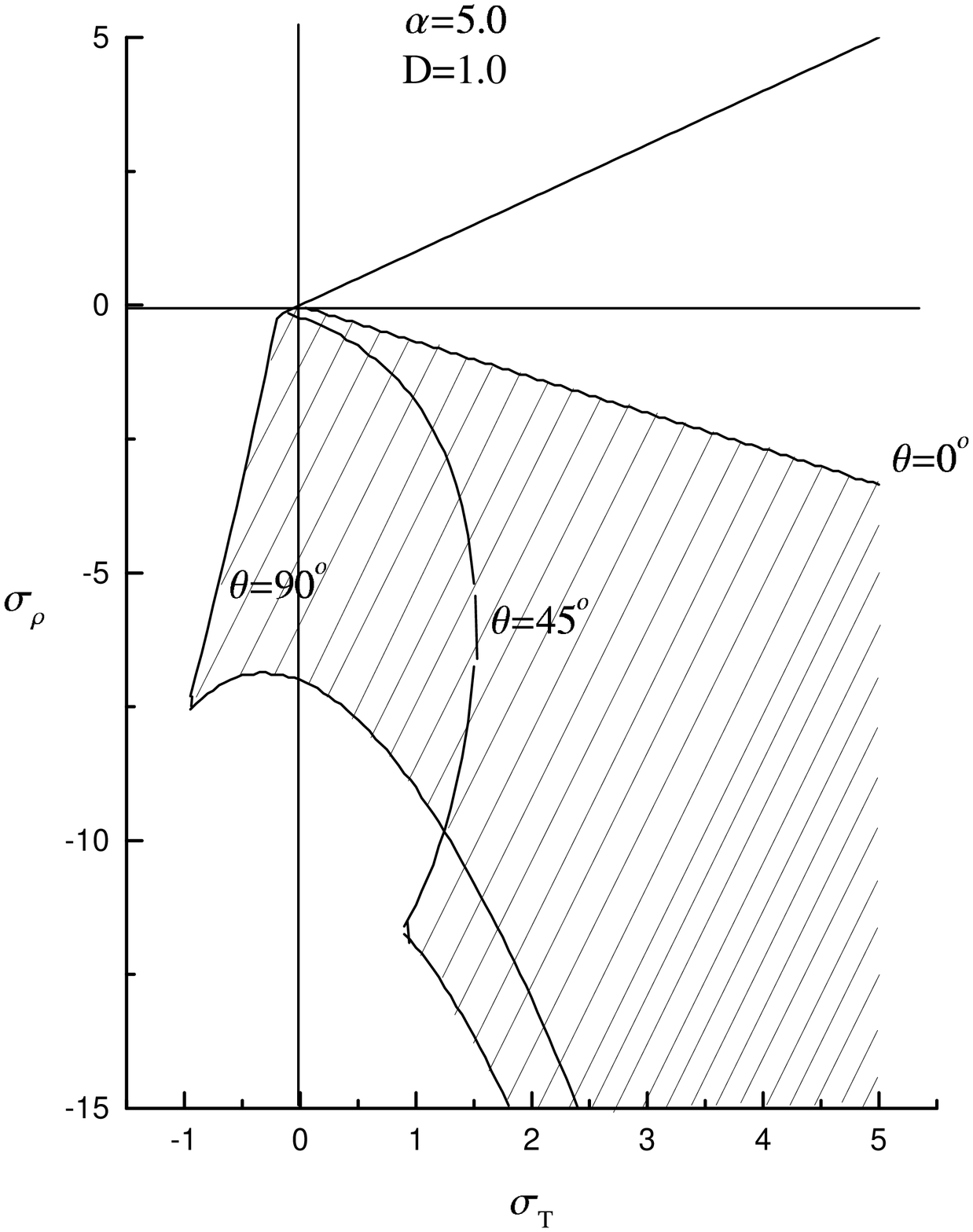}}
   {\epsfxsize=1.7in\epsfysize=2.2in\epsffile{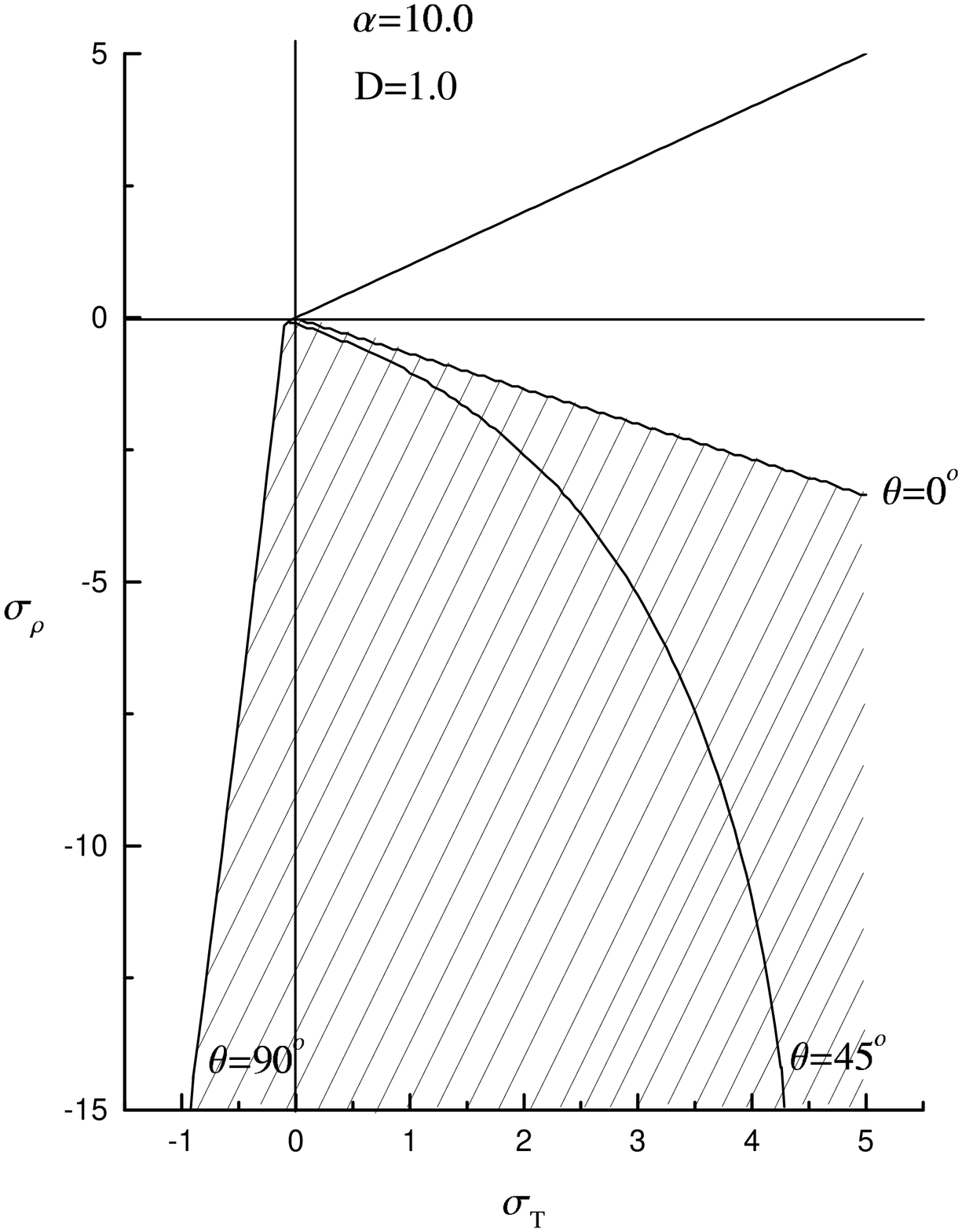}}}
  \centerline{{\epsfxsize=1.7in\epsfysize=2.2in\epsffile{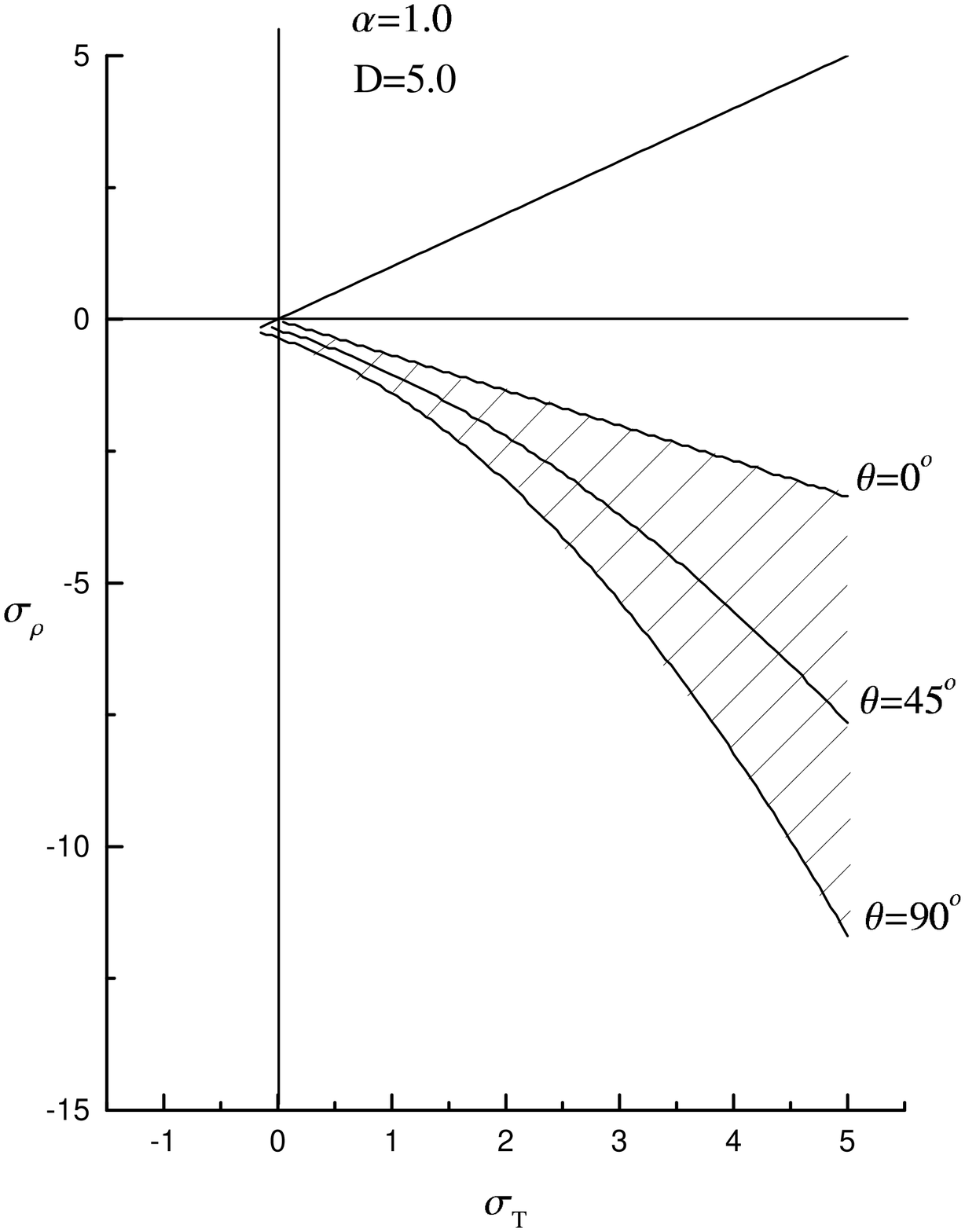}}
   {\epsfxsize=1.7in\epsfysize=2.2in\epsffile{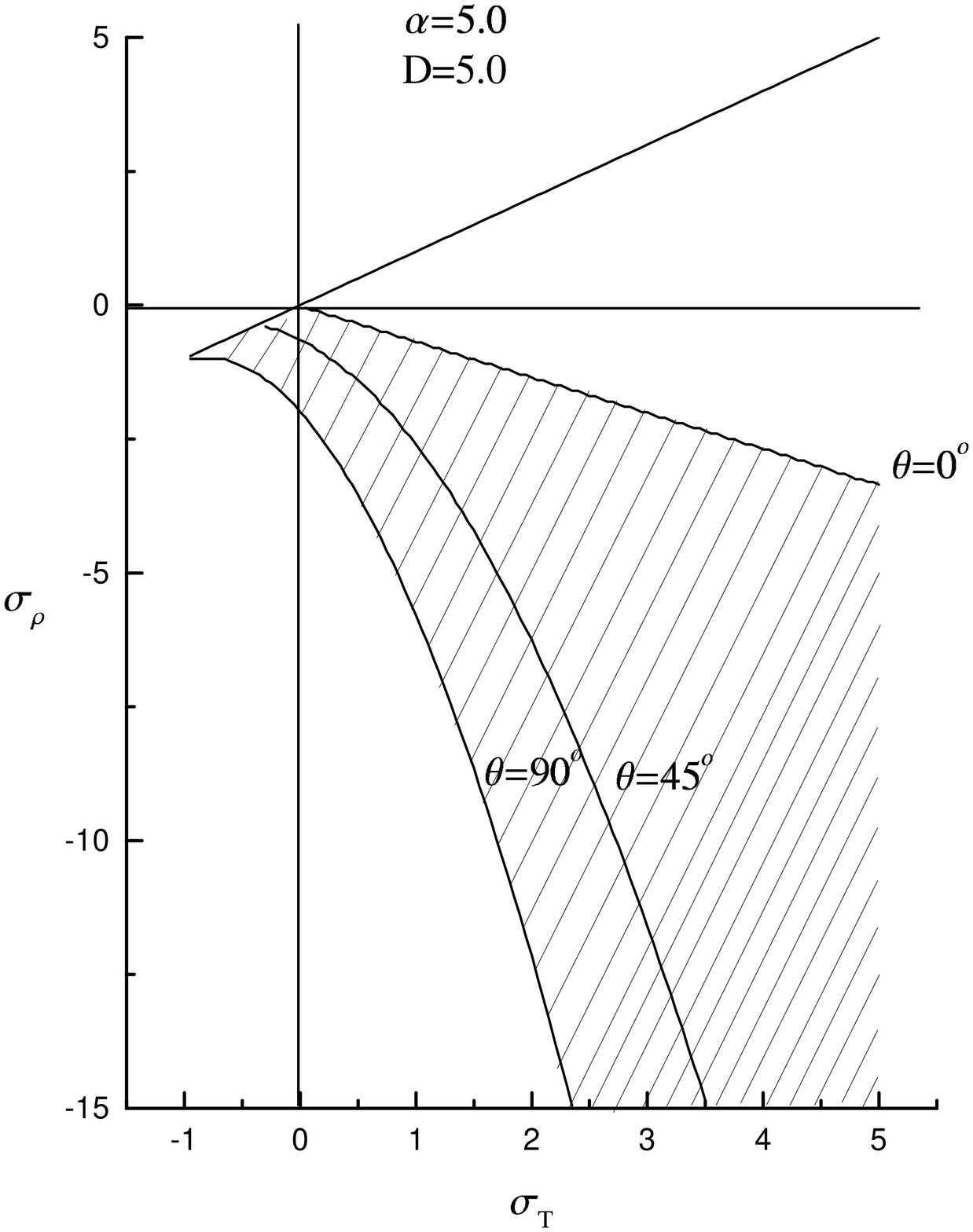}}
   {\epsfxsize=1.7in\epsfysize=2.2in\epsffile{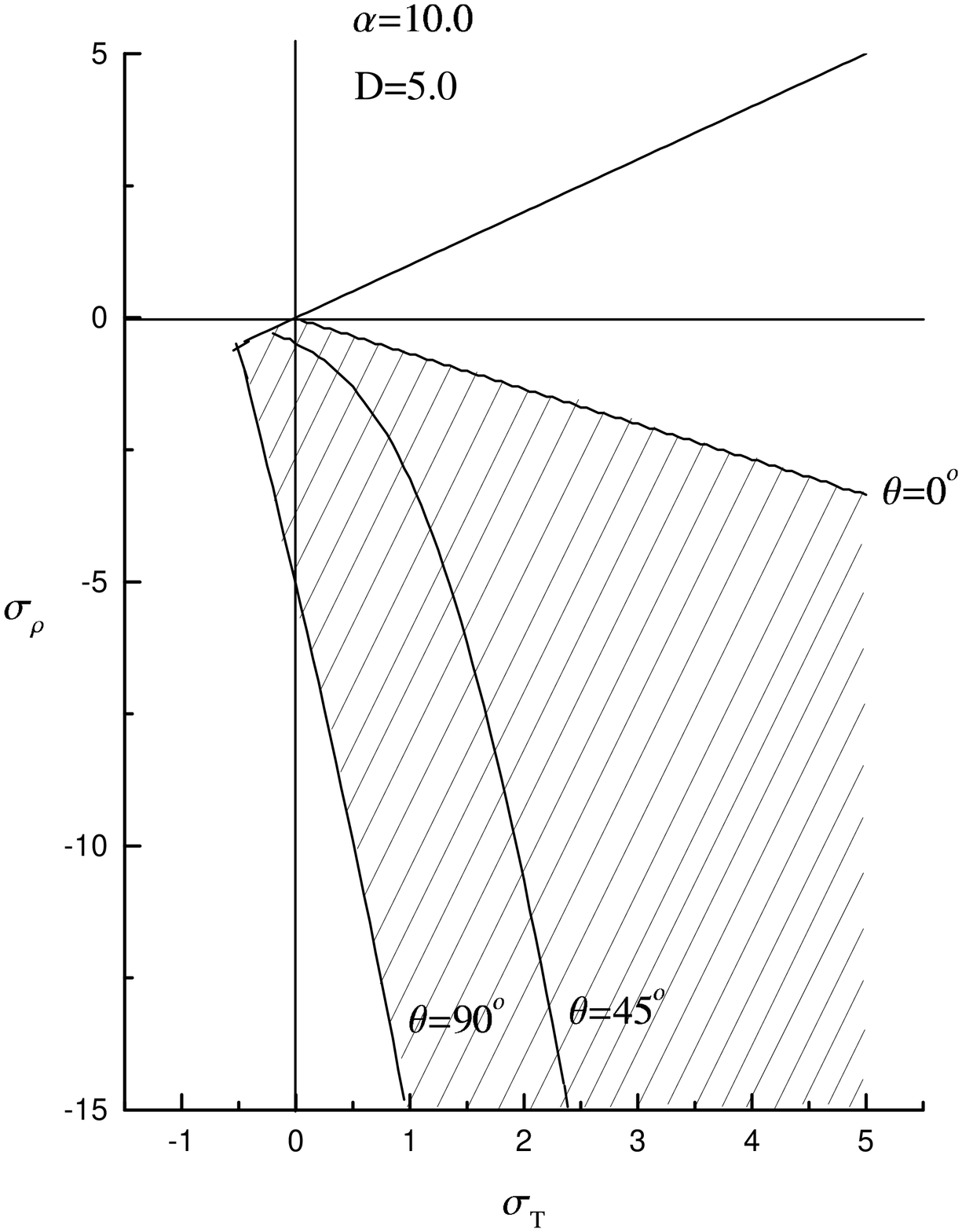}}}
  \centerline{{\epsfxsize=1.7in\epsfysize=2.2in\epsffile{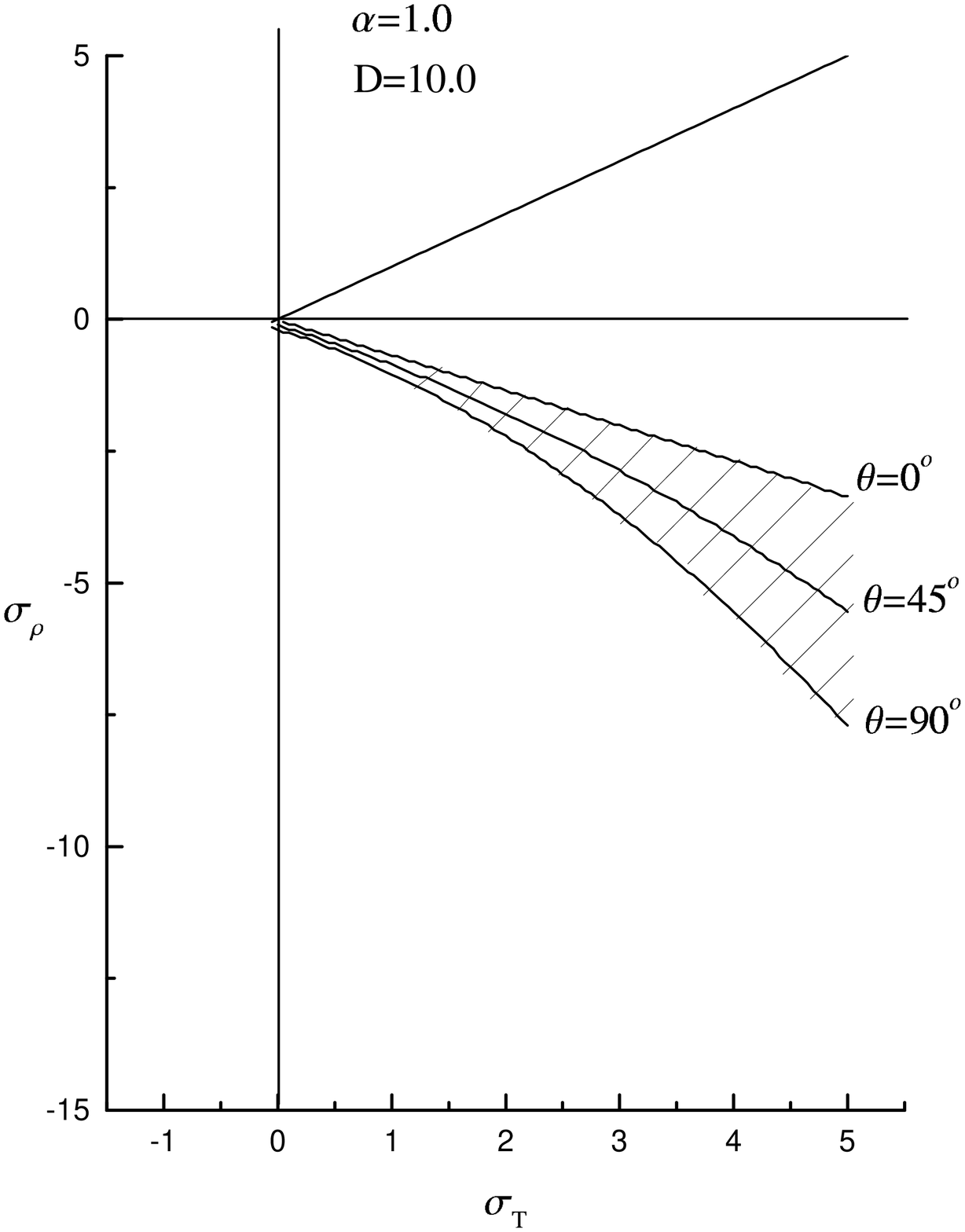}}
  {\epsfxsize=1.7in\epsfysize=2.2in\epsffile{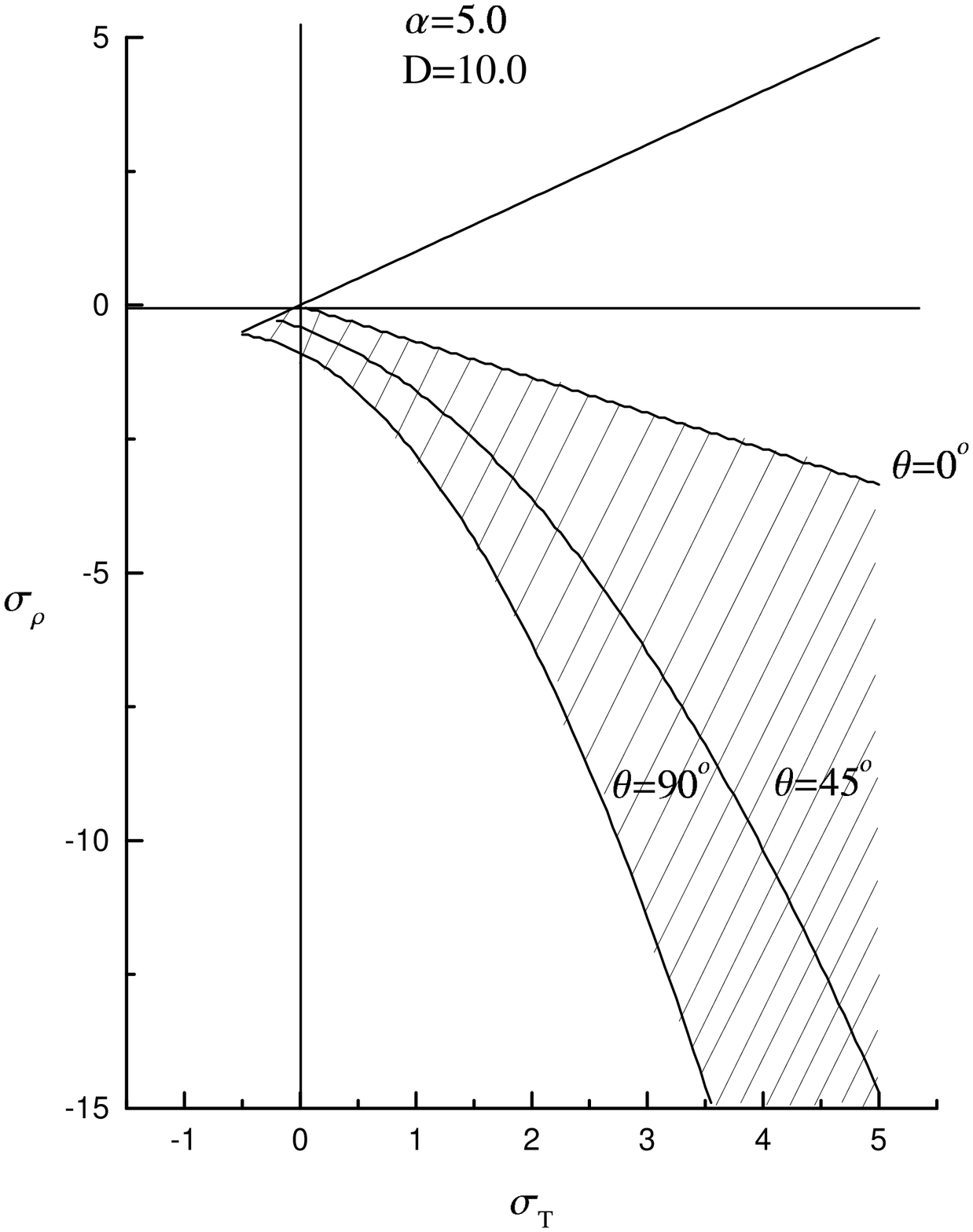}}
  {\epsfxsize=1.7in\epsfysize=2.2in\epsffile{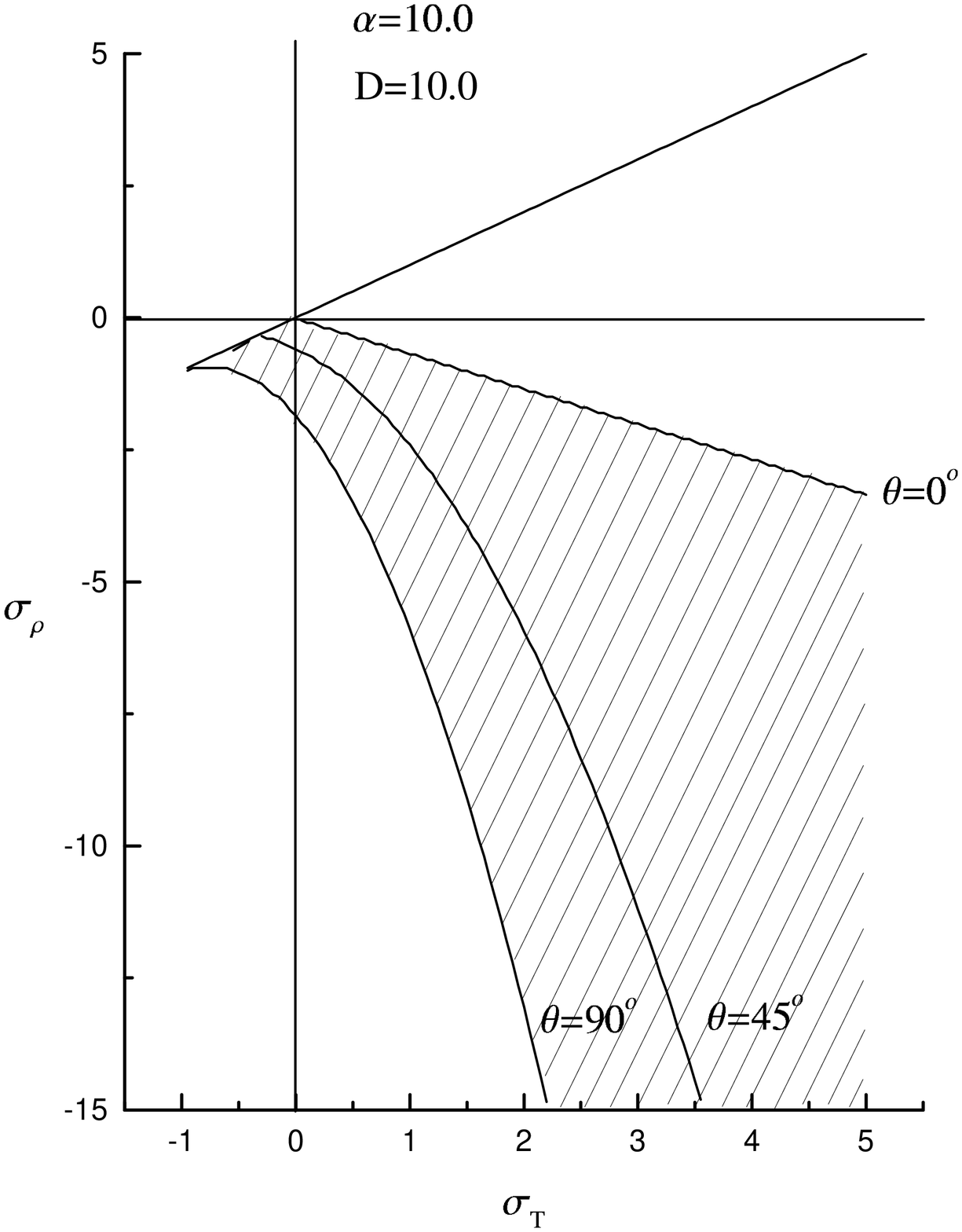}}}
 \caption[]{Regions of stability in the $\sigma_T-\sigma_\rho$ plane
  for presence of magnetic field including ambipolar diffusion
  ($\alpha\neq0,D\neq0$) for three values of $\theta$. The shaded
  areas are semi-stable regions where a disc-like clump can be formed.}
\end{figure}
\newpage
\input{epsf}
\epsfxsize=4in \epsfysize=6in
\begin{figure}
\centerline{\epsffile{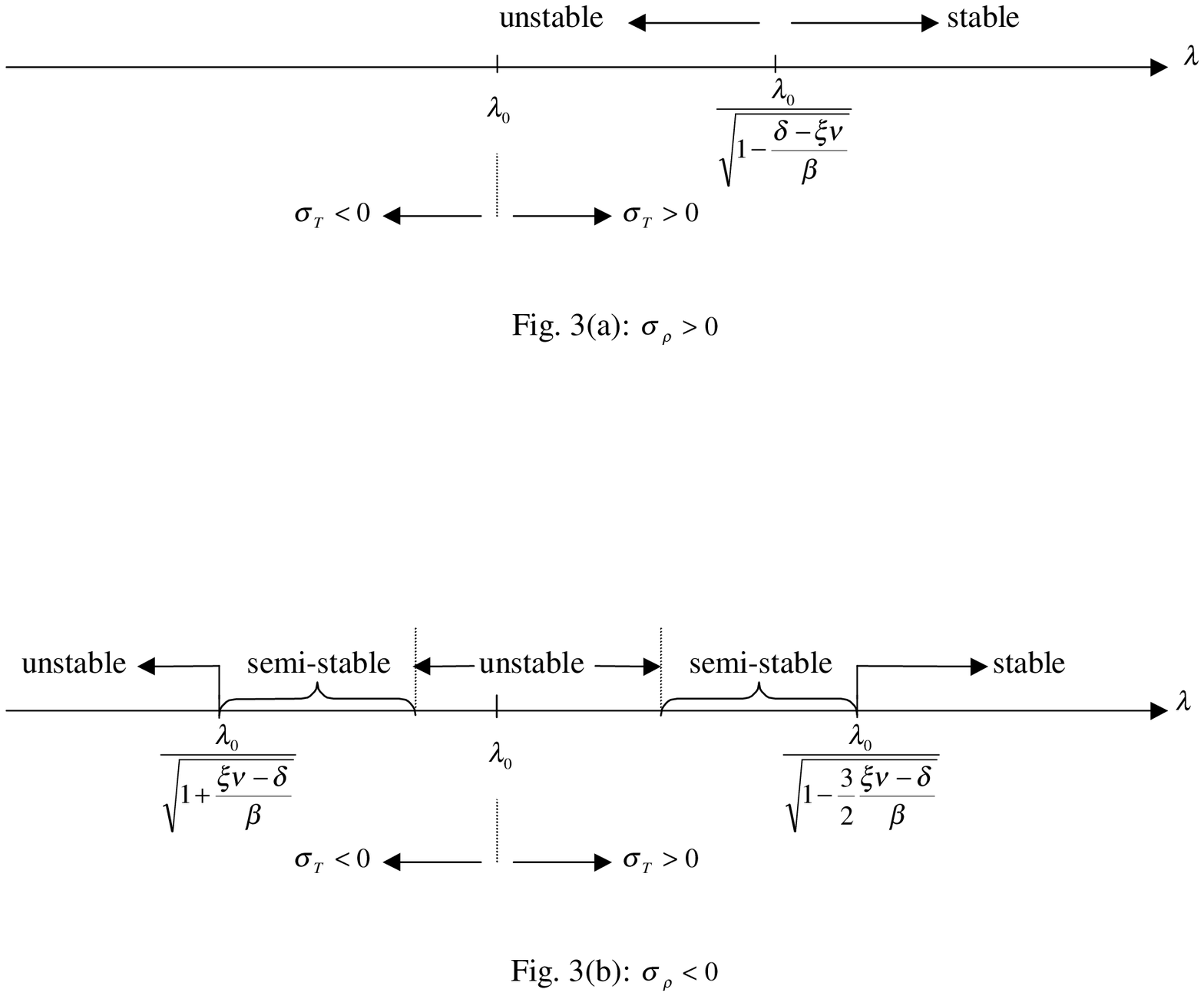}}
  \caption[]{Different cases of stability, unstability, and semi-stability
  of the $\sigma_T-\sigma_\rho$ plane in terms of $\lambda$, wavelength
  of perturbation. Semi-stable regions of figure (b) depend on the magnetic field
($\alpha$) and the ambipolar diffusion strength (D).}
\end{figure}
\newpage
\input{epsf}
\epsfxsize=4in \epsfysize=6in
\begin{figure}
\centerline{\epsffile{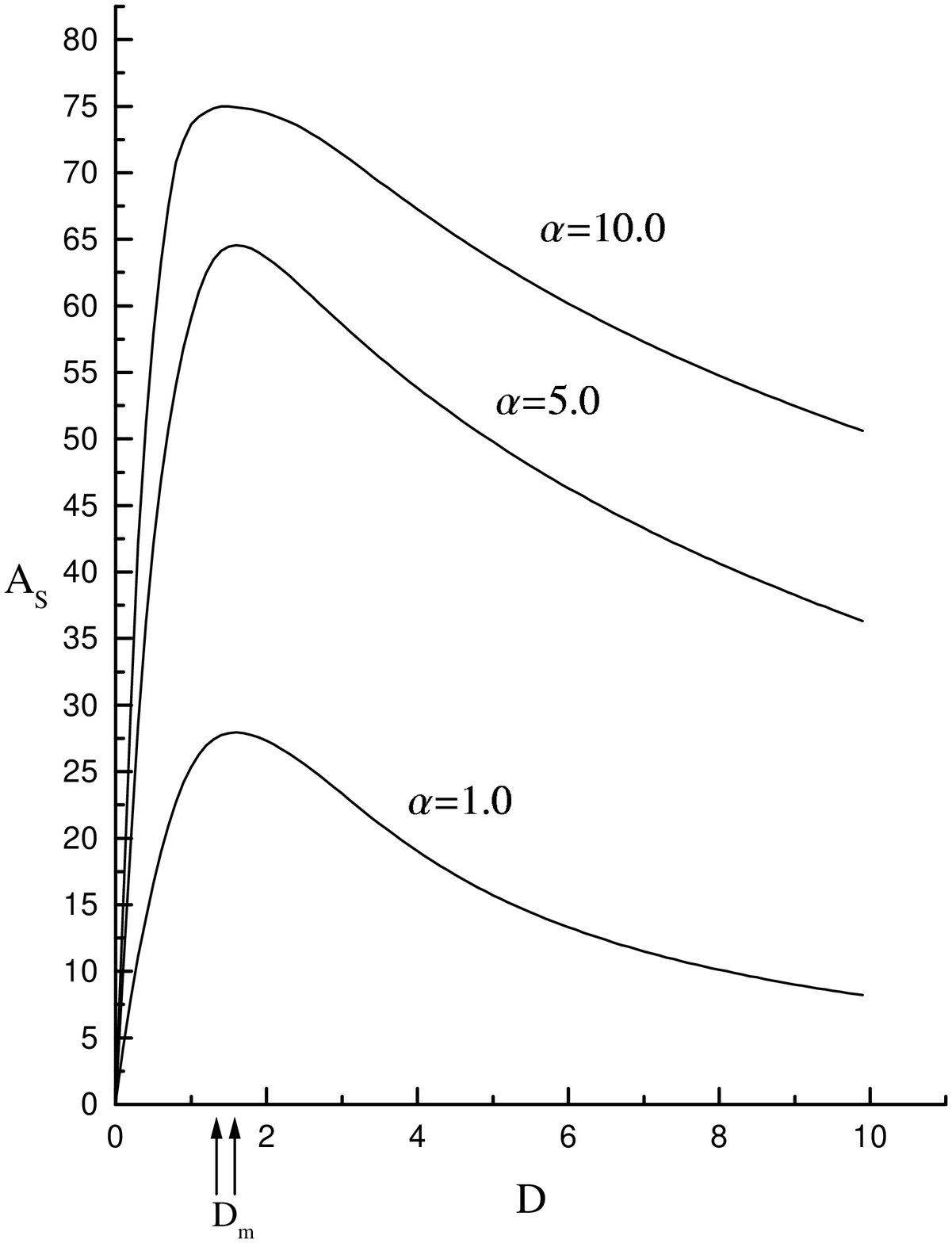}}
  \caption[]{The area of semi-stable region, as a function of $D$ for
  three typical values of $\alpha$. Maximum of semi-stable region
  is occurred at a typical $D_m$.}
\end{figure}
\end{document}